\documentclass{ws-procs961x669}            

\begin{document}
\title{ASTENA: a mission concept for a deep study of the transient gamma-ray sky and for nuclear astrophysics}

\author{E.~Virgilli$^{1,2,*}$, F.~Frontera$^{1,3}$, P.~Rosati$^{1,3,4}$, C.~Guidorzi$^{1,3,4}$, L.~Ferro$^3$, M.~Moita$^3$, M.~Orlandini$^1$, 
F. Fuschino$^{1,2}$, R.~Campana$^{1,2}$,  C. Labanti$^1$, E. Marchesini$^1$, E.~Caroli$^1$, N. Auricchio$^1$, J. B. Stephen$^1$, C.~Ferrari$^5$, S.~Squerzanti$^4$, S.~Del Sordo$^6$, C.~Gargano$^6$ and M.~Pucci$^7$\\ on behalf of the ASTENA collaboration}

\address{
$ $ \newline
$^1$ INAF - OAS di Bologna, Via Piero Gobetti 93/3, I-40129 Bologna, Italy\\
$^2$ INFN Sezione di Bologna Viale C. Berti Pichat, 6/2 – 40127 Bologna, Italy\\
$^3$ Dept. of Physics and Earth Science, Univ. of Ferrara, Via Saragat 1, I-44122, Ferrara, Italy\\
$^4$ Istituto Nazionale di Fisica Nucleare, INFN-Sezione di Ferrara, 44122 Ferrara, Italy\\
$^5$ CNR-IMEM Institute, Parco Area delle Scienze 37/A, 43124 Parma, Italy\\
$^6$ INAF/IASF-Palermo, Via Ugo La Malfa 153, 90146 Palermo, Italy\\
$^7$ CNR - Istituto Nazionale di Ottica - Largo Fermi 6, 50125 Firenze, Italy\newline\newline
$^*$ E-mail: enrico.virgilli@inaf.it}

\begin{abstract}
Gamma-ray astronomy is a branch whose potential has not yet been fully exploited. The observations of elemental and isotopic abundances in supernova (SN) explosions are key probes not only of the stellar structure and evolution but also for understanding the physics that makes Type-Ia SNe as standard candles for the study of the Universe expansion properties. In spite of its crucial role, nuclear astrophysics remains a poorly explored field mainly for the typical emission lines intensity which are vanishing small and requires very high sensitivities of the telescopes.
Furthermore, in spite that the Galactic bulge-dominated intensity of positron annihilation line at 511 keV has been measured, its origin is still a mystery due to the poor angular resolution and insufficient sensitivity of the commonly employed instrumentation in the sub-MeV energy domain.
To answer these scientific issues a jump in sensitivity and angular resolution with respect to the present instrumentation is required. Conceived within the EU project AHEAD, a new high energy mission, capable of tackling the previously mentioned topics, has been proposed. This concept of mission named ASTENA (Advanced Surveyor of Transient Events and Nuclear Astrophysics), includes two instruments: a Wide Field Monitor with Imaging and Spectroscopic (WFM-IS, 2 keV - 20 MeV) capabilities and a Narrow Field Telescope (NFT, 50 - 700 keV). Thanks to the combination of angular resolution, sensitivity and large FoV, ASTENA will be a breakthrough in the hard X and soft gamma--ray energy band, also enabling polarimetry in this energy band. In this talk the science goals of the mission are discussed, the payload configuration is described and expected performances in observing key targets are shown.
\end{abstract}

\begin{figure}[!t]
\centerline{\includegraphics[width=9.5cm]{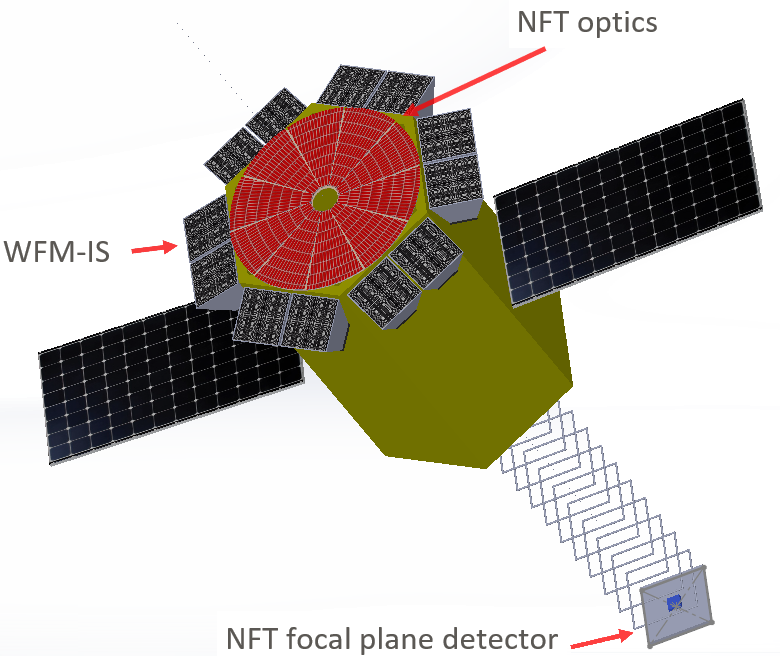}}
\caption{Drawing of the ASTENA mission. The NFT (in red) is a 20~m focal length Laue lens made with bent Germanium and Silicon crystals. At the focal distance 
is positioned a focal plane detector (blue) which is a solid state device made 
with 4 layers of Cadmium Zinc Telluride. The focal distance is achieved 
through a deployable mast which at launch is fully enclosed in the 
spacecraft and in operative condition extends from the bottom of 
the spacecraft for 15~m, which is a reasonable distance for the stability 
of an extendable structure.
The WFM-IS composed of 12 Position Sensitive Detectors (PSDs) 
distributed around the NFT and oriented 15 degrees two by two 
outwards with respect to the Laue lens axis in order to extend 
the FoV of the overall instrument.}
\label{fig:ASTENA}
\end{figure}

\section{Introduction}
\label{intro}
Within the H2020 European program AHEAD\cite{piro2015, piro2017}, (integrated Activities for High Energy Astrophysics Domain) devoted to the assessment of future gamma-ray experiments, a Scientific Advisory Group recommended the prime scientific questions that might be addressed by a future space mission operating in the gamma-ray energy domain.  The high-priorities themes resulted to be the nuclear astrophysics and the study of the transient sky. 
According to those themes, the ASTENA (Advanced Surveyor 
for Transient Events and Nuclear Astrophysics) has been designed. The ASTENA concept mission, which is shown in Fig.~\ref{fig:ASTENA}, is a broad energy pass-band experiment composed by two complementary instruments. The first is a Wide Field Monitor with Imaging and Spectrometric capabilities (WFM-IS) with an outstanding broad energy pass-band from 2~keV to 20~MeV. It consists on 12 coded mask cameras deployed in a circular pattern around the hexagonal spacecraft and oriented at 15~degrees with respect to the spacecraft axis. The overall Field of View (FoV) of the instrument is $\sim$2~steradians. The second instrument is a narrow FoV (few arcminutes) telescope based on a Laue lens with a geometric area of $\sim$7~m$^2$ and 20~m focal length, capable to focus photons in the broad energy pass-band 50 - 700~keV  on a solid state detector. The Narrow Field Telescope (NFT) represents a technological breakthrough as its optics, based on the diffraction from bent crystals, provides an unprecedented sensitivity with respect to any other mission flown and operative in the same energy pass-band. At launch, we expect to keep the WFM-IS cameras and the focal 
plane detector stored within the cylindrical spacecraft whose
diameter is 1.5~m  and the length is 5~m. In the 
operational configuration an extendable boom brings the focal 
plane detector 15~m apart from the bottom of the spacecraft while 
a mechanism discloses and tilts the coded mask cameras at their nominal position and angle. In the following currently on-going project called AHEAD2020 the mission concept is being refined and optimized. The mission has been proposed in the ESA call "Voyage 2050"\cite{favata21} as a future medium class mission for hard X-/soft gamma-ray 
astrophysics\cite{frontera19wp, guidorzi19wp}.  The final recommendation of the Senior Committee has confirmed the key importance of the high energy observations from space 
with high sensitivity and capable of enabling spectro-polarimetry based on new 
technologies, particularly in synergy with gravitational wave astronomy for resolving 
some of the fundamental questions still unanswered in astrophysics related to the nucleosynthesis in explosive events or to the accretion mechanism on compact sources.

\begin{figure}[!t]
\includegraphics[width=6.3cm]{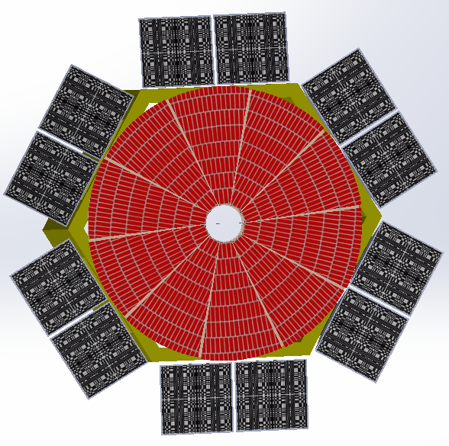}
\includegraphics[width=6.3cm]{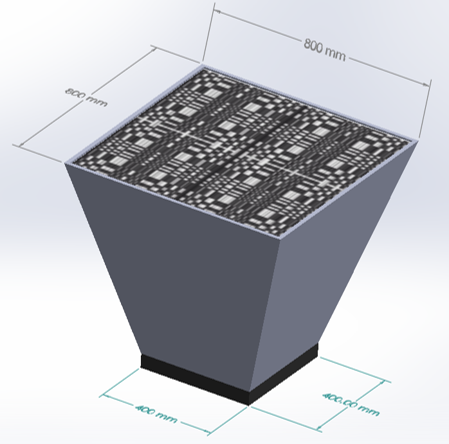}
\caption{Left: Top view of the instruments on board ASTENA. The 12 PSD of the 
WFM/IS are coupled 2 by two and are positioned 
around the hexagonal structure of the spacecraft. Each PSD is surmounted by 
a coded mask which ensures the imaging capability up to 150~keV. 
The red part is the broad energy pass-band Laue with 15 m focal length. 
Right: Detail of one PSD in which is visible the coded mask and the collimator surrounding the instrument. At the bottom (dark profile) is visible the detection plane made with scintillator bars which are coupled with SDDs (see text for the explanation of the instrument working principle).}
\label{fig:ASTENAdetails}
\end{figure}

\section{The ASTENA configuration}
\label{configuration}

The ASTENA mission concept builds on the complementarity between the two 
instruments on board. The WFM-IS is composed by 12 coded mask cameras 
equipped with Position Sensitive Detectors (visible in 
Fig.\ref{fig:ASTENAdetails} - left)
(PSDs) distributed two by two around the hexagonal 
spacecraft envelope. 
Each pair of PSDs in the same hexagon side is 
co-aligned and directed towards the same direction 
which is radially tilted outwards of 15 degrees with 
respect to the satellite axis.
Each PSD has a size of 43 $\times$ 42 cm$^2$ 
and is made of $\sim$
6500 scintillator bars with hexagonal cross section (5 mm between 
flat sides) and 50 mm 
long. Both ends of the scintillator bars are 
optically coupled with two Silicon Drift Detectors (SDDs) 400 $\mu$m thick. 
The instrument is based on the same detection 
principle of the X and Gamma-ray Imager and Spectrometer 
(XGIS) on board the THESEUS mission\cite{amati2017} which was proposed as 5$^{th}$ medium-class mission for the ESA Cosmic Vision Programme (M5).

\begin{figure}[!t]
\includegraphics[width=6.3cm]{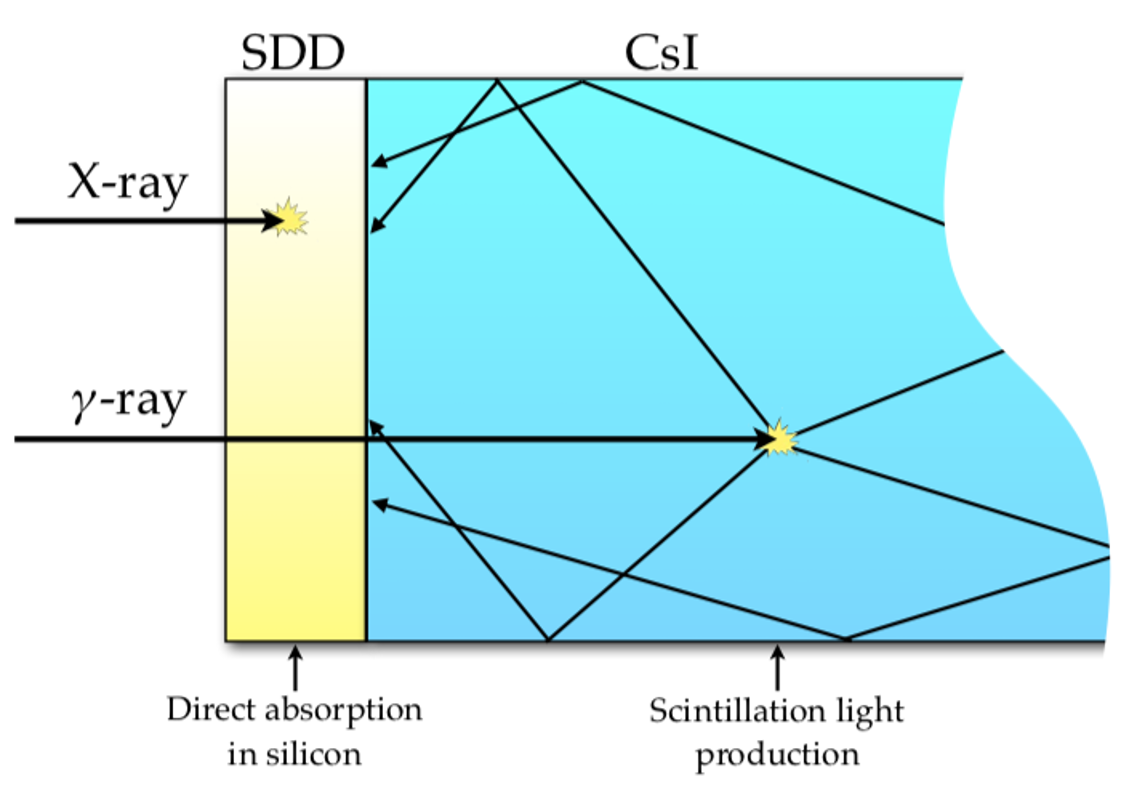}
\includegraphics[width=6.3cm]{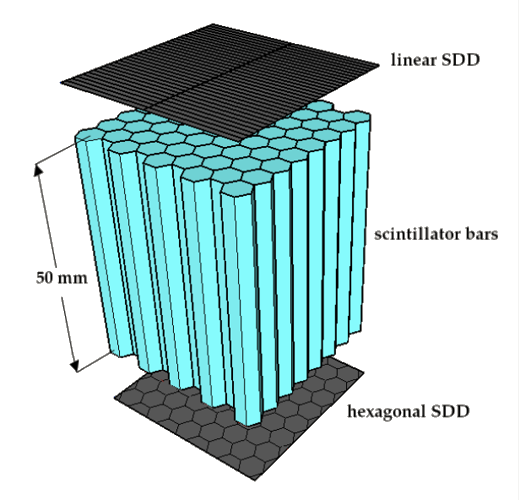}
\caption{Left: working principle of the WFM-IS. An hexagonal scintillator bar (cyan) is coupled at both ends (only one end is visible in the picture) with an SDD (yellow). The low energy radiation ($<$ 30~keV) is directly absorbed in the SDD while higher energy photons pass through the SDD and interact in the scintillator bar. The scintillation light, which is produced at some distance from the surface of the scintillator bar and is reflected by the lateral surfaces, is collected by the two SDDs. The comparison between the two signals allows for the determination of the depth of interaction. Right: disposition of the scintillator bars (cyan) and of the top and bottom SDDs (grey). The top SDD is equipped with an array of linear anodes therefore the charge collected provides information only in one direction. In the bottom SDD the anodic structure has a hexagonal shape in order to fully exploit this geometry which minimizes the directional bias for polarimetric measurements, if compared with a detector with square (cubic) pixelization (voxelization).}
\label{fig:sdd_details}
\end{figure}

\begin{figure}[!t]
\includegraphics[width=6.3cm]{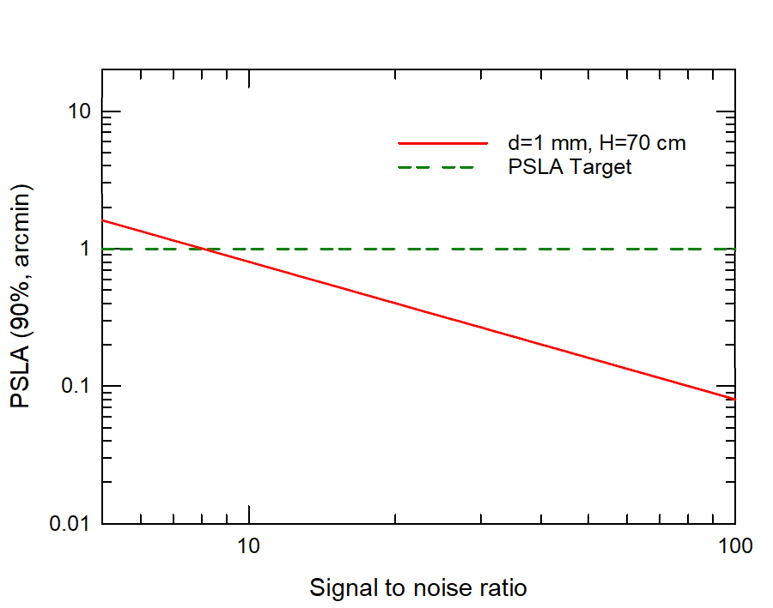}
\includegraphics[width=6.3cm]{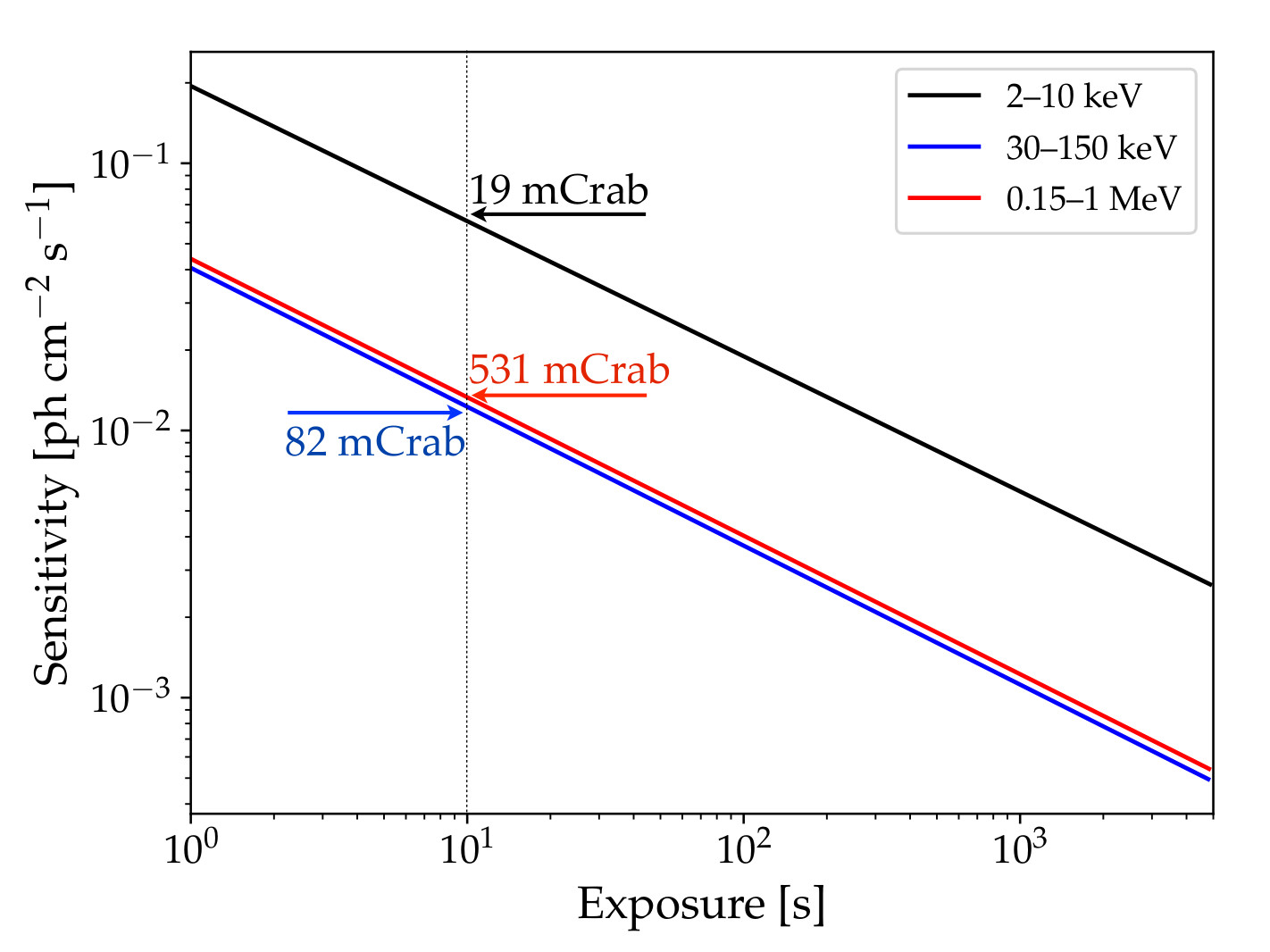}
\caption{Left: Point Source Location Accuracy (PSLA) for 
the WFM-IS as a function of the signal to noise ratio. 
The parameter d is the distance between the anodes  of 
the top SDDs. The requird PSLA is also   shown  (green 
dashed line) Right: Integrated continuum   sensitivity 
expressed in photons/cm$^2$/s as a function  of    the 
exposure time for the three significant energy pass-bands 
acquired for direct absorption 
in the SDDs (black curve, 2 - 10~keV), through the 
scintillation bars (30 - 150~keV, blue curve) and as 
a Compton detector with no coded mask (150~keV - 
1~MeV, red curve).}
\label{fig:wfm_performances}
\end{figure}

The detection principle (see Fig.~\ref{fig:sdd_details})
is based on a different interaction of the radiation with 
the system, depending on the energy of the interacting 
photons. Low energy photons ($<$ 30~keV) are directly 
absorbed in the SDD while photons with higher energy pass 
through the SDD and interact in the scintillator bar. 
The scintillation light, which is produced at some distance 
from the ends of the scintillator bars, is reflected by 
the lateral surfaces which are properly polished and wrapped 
in order to reflect as much as possible the light produced,  
and it is collected by the two SDDs optically coupled at both 
ends of the bar. For high energy photons ($>$30~keV) the detector 
is position sensitive with a 3-D resolution
given that the position of interaction of 
the gamma-ray photon along the bar can be reconstructed
by comparing the signal collected from the two opposed 
SDDs. 
The top SDDs (those facing the sky) have the dual 
purpose of directly detecting the low energy photons 
(below few tens of keV) and of collecting the 
scintillation light emitted by the scintillator bar 
when photons with higher energy pass through them. 
They have a linear anodic structure therefore are
sensitive only to one direction (X or Y) with
spatial resolution of 1.25~mm. Given that each pair 
of PSDs placed at the same side of the spacecraft 
have perpendicular anodes, they behave 
like a 2-D detector. Instead, the bottom SDD has 
hexagonal shape and it is used to get the 3-D 
position sensitivity. 
In this way we can minimize the background through
the Compton kinematics reconstruction of the 
trajectory of the photons and exploit the polarimetric 
capability of the instrument. 
Different scintillator 
materials as Cesium Iodide (CsI(Tl)), 
Gadolinium Aluminium Gallium Garnet (GAGG), 
Lutetium Yttrium Orthosilicate (LYSO(Ce)) 
are under investigation in order 
to find the optimal 
properties to fit the instrument requirements.
Each PSD is surmounted by a double scale\cite{skinner93} 
square coded mask with side of 80~cm at a distance of 70~cm from 
the detector plane (see Fig.~\ref{fig:ASTENAdetails} - right). 
The double scale enables imaging with Point Source Location 
Accuracy (PSLA) of 1~arcmin for a 7$\sigma$~signal 
for photons with energy $<$ 30~keV and PSLA of 
5~arcmin for photons in 
the energy range 30 - 150~keV.
In Fig.~\ref{fig:wfm_performances} (left) the PSLA as a function of the significance of the observation for different configurations of the WFM-IS is reported, compared 
with required PSLA of 1~arcminute. 
In Fig.~\ref{fig:wfm_performances} (right) the integrated sensitivity of the overall WFM-IS is reported as a function of the integration time, divided in three relevant pass-bands according to the different detection principle employed (2 - 10~keV: direct detection in the SDDs, 30 - 150~keV: interaction with the scintillator bars, 150~keV - 10~MeV: uncollimated interaction in the scintillator bar).
Above 150~keV the coded mask is transparent to radiation 
and an effective imaging capability  
can be enabled through two features: 
1. by exploiting the Compton kinematics for 
the photon direction reconstruction and 2. by taking advantage of
the different measured photon intensity from the 6 blocks 
which are - in general - differently oriented with respect to 
the direction of the observed event, except for perfectly 
on-axis sources. With these configuration the instrument can 
provide a FoV of $\sim$2~sr with an angular 
resolution of a few arcmin and an unprecedented energy 
pass-band of 2~keV - 20~MeV which has never been 
achieved so far with a single device.

The second instrument is a 20~m long focal length Laue lens based on 
bent crystals made with Germanium and Silicon. The crystals are 
distributed in 43 concentric rings and, according to the 
spherical geometry and to the Bragg law, the lower energies 
are reflected from outer rings while inner rings are responsible 
for focusing the higher energies.
The size of the crystals have been chosen to be 30 
$\times$ 10 mm$^2$, the longer dimension being the focusing direction. 
In the other direction, no concentration is expected due to the 
cylindrical curvature of the tiles. The cross section of the tiles
has been chosen in order to tile the overall geometric area with a 
moderate number of crystals ($\sim$19500) by minimizing the amount 
of uncovered area (the optics filling factor results above 85\%).
Depending on the energy to be diffracted the crystal thickness is 
optimized in order to maximize the diffraction efficiency. For both 
materials the 111 planes are assumed in order to exploit the so 
called secondary curvature of the diffraction planes 
induced in crystals for some crystallographic 
orientation (including the 111) allowing to achieve a throughput 
which overcomes the limitation of the 50\% of the incident beam, which 
is the limit of mosaic and flat perfect crystals.

\begin{figure}[!t]
\includegraphics[width=12cm]{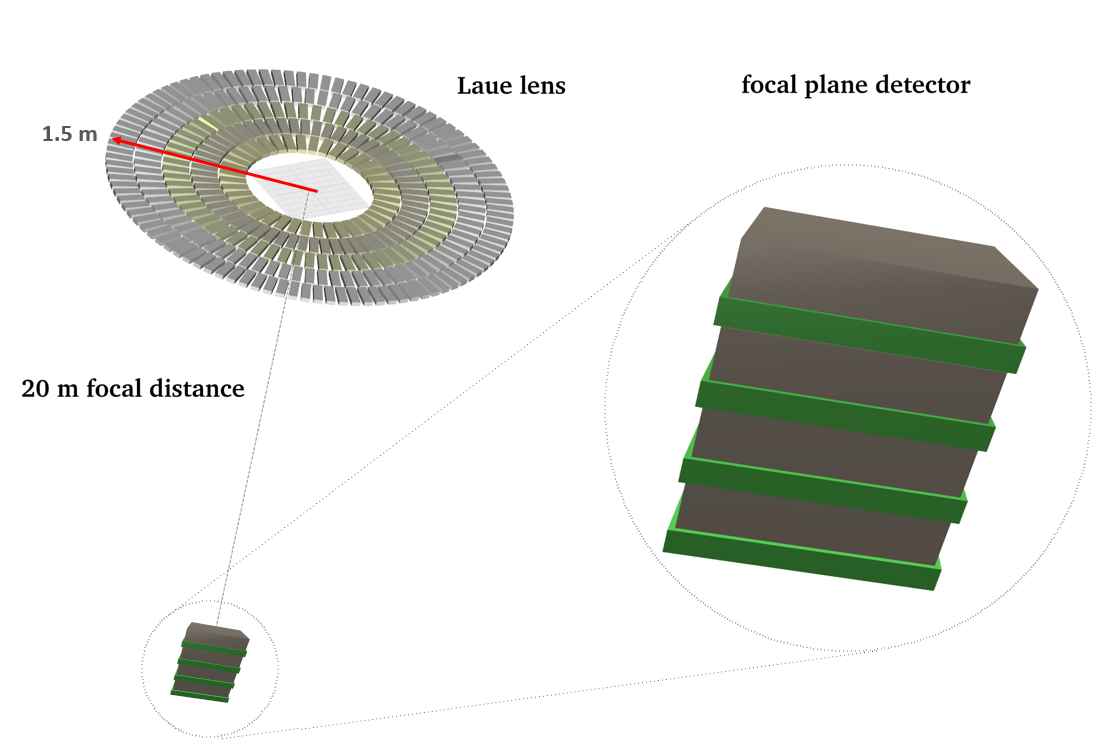}
\caption{Left: The optics of the NFT composed by 43 rings of crystal 
tiles made with Silicon and Germanium. The optics has a radius of 
1.5~m for a total geometric area of about 7~m$^2$. At 20~m from the Laue
lens is placed the focal plane detector. Right: a Geant4 drawing 
of the main components of the focal plane detector which is made 
of 4 layers of a solid state Cadmium Zinc Telluride (CZT, dark grey) 
with 3-D spatial resolution, spectral and polarimetric capabilities.}
\label{fig:nft}
\end{figure}

Orders of diffraction higher than the fundamental are being considered 
in the simulations in order to maximize the effective area, particularly 
in the energy range 500 - 520~keV (at the expenses of the effective area 
at lower energies) in order to increase the sensitivity at the energy 
of interest for the detection of the weak e$^+$/e$^-$ annihilation line
from the Galactic center (see Sect.~\ref{science} for further details).
At 20~m from the optics is placed the focal plane detector which is kept at the correct position with a deployable mast. The detector is made with 4 layers of Cadmium Zinc Telluride (CZT) with cross section 80 $\times$ 80 mm$^2$ each with a thickness of 20 mm. The detector has a 3-D spatial resolution of about 300 $\mu$m in all directions. This is achievable with a proper disposition of the anodes on the top and bottom of each CZT layer. Thanks to the overall detector thickness its detection efficiency is greater than 80\% at 600~keV 
with energy resolution of 1\% at 511~keV. 

\begin{figure}[!t]
\centerline{\includegraphics[width=15cm]{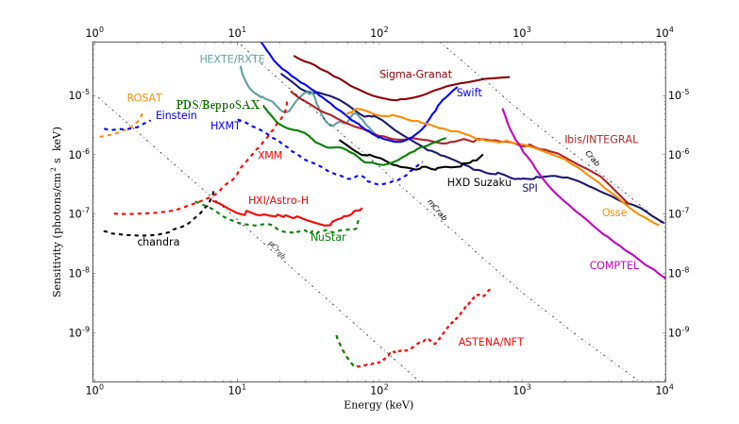}}
\centerline{\includegraphics[width=9.5cm]{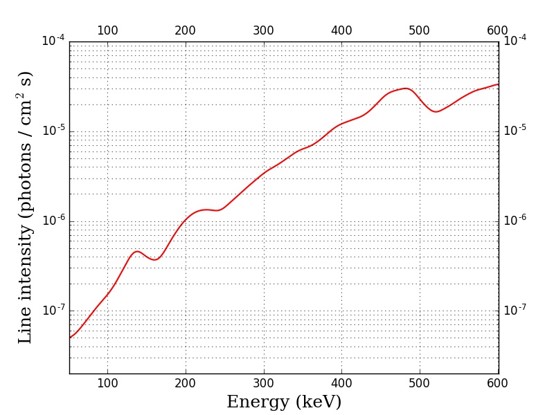}}
\caption{Top: NFT continuum sensitivity at 3$\sigma$ confidence 
level and in 10$^5$~s of observation time, with $\Delta$E = E/2, 
except in the band 50 to 62.5~keV, where $\Delta$E linearly 
increases from $\Delta$E = E/4 to $\Delta$E = E/2, compared 
with the sensitivity at the same significance and with the same 
integration time of other missions or experiments (dashed lines 
represent other focusing experiment, continuum curves 
correspond to direct view instruments). Bottom: Expected 
line sensitivity for the NFT, calculated for an integration 
time of 10$^5$~s, at 3$\sigma$ confidence level.}
\label{fig:nft_sensitivity}
\end{figure}

The NFT continuum sensitivity which has been derived at Low Earth Orbit
(LEO) by comparison with the background measured by the Spectrometer on-board INTEGRAL SPI\cite{verdrenne2003} is reported in Fig.~\ref{fig:nft_sensitivity} (top). The curve has been estimated for an integration time of $10^5$~s 
and at 3$\sigma$ confidence level. 
For comparison, in the plot has been reported a number of 
past and present experiments exploiting the direct view principle 
(equipped with coded masks or collimators) or using focusing 
telescopes. The energy band in which it is estimated is 
$\Delta\,E = E$/2, except in the band from 50~keV to 62.5~keV, 
in which the energy band linearly increases from 
$\Delta\,E = E$/4 to $\Delta~E = E$/2, due to 
the absence of crystals diffracting energies below 50~keV.
Uncertainties in the realization of the optics have been also taken 
into account. The curvature radius of the crystals have been considered within 5\% of the nominal radius and the mounting accuracy of 
about 10~arcseconds with respect to the nominal positioning. 
The presence of such deviations from the nominal lens 
realization requirements have effects on the size of the PSF and, 
ultimately, on the focusing power of the Laue lens.
This unprecedented sensitivity is achieved  thanks to the use 
of bent crystals with optimized thickness.
Figure~\ref{fig:nft_sensitivity} (bottom) shows the sensitivity 
at measuring the flux of emission lines, at $3\sigma$ confidence 
level and for 10$^5$~s integration time. An intrinsic 
FWHM of 2~keV has been considered for the lines (e.g. see 
[\!\!\citenum{diehl14}] for the 158~keV line from SN2014J).
The narrow line sensitivity is about 1 order of magnitude
better than SPI on board INTEGRAL at 511~keV.
These values of sensitivities are mainly due to the use of bent 
crystals, to the transmission geometry and to the focal length that 
allows for an unprecedented large collecting area.

Laue optics have the great advantage of drastically reducing the 
instrumental background as the photons are mainly collected into an 
area of a few mm$^2$. The drawback is that, as far as the source 
 moves out of the focal axis, the diffracted image spreads over a 
 larger area generating ring-shaped images for coma aberrations. 
 Through ray-tracing and Monte Carlo simulations it has been 
 estimated that, with the defined optics dimension and 
 crystal size, photons from off-axis sources ($>$4~arcmin)
 are scattered into a broad detection area\cite{frontera21}. 
 This reduces the benefits of the focusing process. Therefore 
 the FOV of the NFT has been fitted with the PSLA of the WFM-IS 
 in order to  exploit the sensitivity 
 of the former for nearly on-axis sources with 
 the excellent broad-band survey capabilities of 
 the latter.

\section{Key science with ASTENA}
\label{science}

For some relevant scientific key subjects that 
can be tackled with the instruments on board ASTENA see 
[\!\!\citenum{frontera21, guidorzi21}]. Here we 
summarize these subjects. As mentioned in Sect.\ref{intro} 
the main objectives of the ASTENA mission concept are mainly 
two. First, ASTENA intends to investigate the intriguing key 
questions related to the gamma-ray lines in astrophysics. 
Furthermore, the high sensitivity of the on board 
instruments joined with the 
large pass-band of the WFM-IS and their polarimetric 
capabilities would make ASTENA an ideal tool for 
shading light in the transient sky, providing also an 
important contributions to the multi-messenger synergy.

\subsection{The transient sky observed with ASTENA}

The discovery of the Gravitational Wave (GW) event 
GW170817\cite{abbott2017} and its
electromagnetic counterpart GRB170817A\cite{kienlin2017, 
connaughton2017, goldstein2017, savchenko2017} marked
the birth of the multi-messenger study of the transient 
sky. In the near future, this plural effort will solve 
fundamental astrophysics and cosmological questions. 
Furthermore, it will boost the discovery of sources 
of known and unknown classes 
of transients. Short GRBs represent the main class 
of transient which has been already confirmed to be 
associated with powerful GW events. 

%
%

Polarization level in the range 30 - 80\% 
of the prompt emission of the GRBs has been 
claimed for a few dozens events (a summary 
of the properties of the events can be found 
in~[\!\!\!\citenum{mcconnell17, gill21}])
but, due to the limited statistical
significance of the results, 
their confirmation is not definitive. 
These uncertainties are mainly 
due to the lack of sensitivity of the 
instruments used as gamma-ray polarimeters. 
A significant detection could probe the geometry 
of the  magnetic field and its intensity which 
are precious information for shedding light into 
the jet composition and dissipation mechanism.
Thanks to its polarimetric capabilities, 
along with  a large detection area 
and broad pass-band, the WFM-IS would be  
the  ideal   instrument  to measure the 
degree   of polarization  of the 
prompt emission,   to   perform a   detailed 
time-resolved  study   and to   evaluate the 
dependence of the polarization degree with energy.

%
%

Instruments capable of measuring the electromagnetic 
counterpart of GW events have a crucial importance 
for different reasons. 
Firstly, they can independently confirm the astrophysical 
nature of a GW trigger, particularly for the faintest events. 
Furthermore, since present interferometers have large 
uncertainties in the sky localization, a high sensitive 
wide FoV instrument like the WFM-IS could provide the 
localization of the order of 1~arcmin with the possibility 
of performing followup observation with a soft 
gamma-ray telescope with an unprecedented sensitivity 
and angular resolution as the NFT.

It must be mentioned that with the launch 
of the Laser Interferometer Space Antenna (LISA) 
a plethora of GW events from stellar mass BH 
binaries will be detected in the mHz regime with a sky 
localization of the order of 1~deg$^2$. Such detections
will work as alert for the observation of the same events
at higher frequencies weeks/months later (with an accuracy 
of the order of tens of seconds) from ground 
interferometers. Thanks to the large FoV and to the PSLA 
of the WFM-IS will be possible to point in advance the 
instrument for the detection of the prompt 
hard X-ray counterpart of the event as well as the
detection of the delayed hard X-ray emission through 
the NFT.

%
%
After the discovery of the GRB afterglow, till 
now this emission has been observed  almost in 
all energy bands from IR to the soft X-ray regime 
up to $\sim$ 10~keV. Afterglow detection in the 
0.1–10~GeV, in the sub-TeV (100–440~GeV) and in 
the TeV energy regions have been reported
\cite{abdalla19, magic19}.
One of the most important open issues that have
still to be settled is the afterglow emission in the 
sub-MeV/MeV region which is almost completely 
undetected except for 
some events detected above few tens of keV 
\cite{maiorano05, kouveliotou13, Martin-Carrillo14}
This is mainly due to the low 
flux emitted in this energy band, joined 
with the lack of sensitivity for the 
present instrumentation. From the prompt 
emission detected with the WFM-IS and 
though a fast repointing it 
would be possible, thanks to the sensitivity 
of the NFT, to measure the hard x-ray 
afterglow as well as its polarization level. 


\subsection{Nuclear astrophysics with ASTENA}

In spite of its importance for understanding the 
inner regions of the astrophysical sources emitting 
hard X and mainly gamma-rays, nuclear astrophysics 
is still  a theoretical field and almost experimentally 
unexplored,  mainly due to observational limits. 
The reason is that present instrumentation in the hard-X 
and gamma-ray regime has absent or crude imaging capability 
and low sensitivity, compared with instruments in other 
wavelengths.
One of the most relevant open issues in astrophysics 
is the origin of the 511~keV positron annihilation line 
observed from the Galactic Center (GC).
Discovered in the seventies\cite{johnson73} 
neither the origin of the gamma emission, nor 
the responsible for the positron production 
have been found. The 511~keV 
annihilation line still represents
a puzzle mainly due to the limitation 
of both sensitivity and angular resolution 
of current instruments operative in the
0.5 MeV region. To date, the best mapping 
of the line was obtained through SPI on-board 
INTEGRAL\citep{knodlseder2007}.
A variety of potential astrophysical sources 
responsible of the 511~keV signal and/or of the 
production of the positrons 
have been proposed, including type Ia supernovae\cite{kalemci2006},
GRBs\cite{bertone2006}, microquasars\cite{guessoum2006}, 
low-mass X-ray binaries\cite{prantzos2004}, 
and neutron star mergers\cite{bartels2018}. 
The high number of X–ray sources 
in the GC suggests the possibility that 
the 511~keV line is due to the emission 
of a number of discrete - unresolved - 
sources. The detection of a transient 0.5~MeV 
emission from V404 Cygni\cite{siegert16-v404} 
is in favour of this hypothesis.
It is also believed that the 511~keV observed 
map would only represent the annihilation 
sites and not the positron sources. 
The propagation of the positrons away from 
the source could be the cause of a general 
broadening which originates the diffuse 
511~keV emission. 
No point sources of annihilation radiation 
have yet been detected in the GC. Nevertheless, 
the angular resolution of INTEGRAL/SPI  does 
not provide any definitive information 
about structure in the emission, even 
with several years of integrated 
data\cite{bouchet10}. Under these considerations 
it is clear that observations of the 
GC with much higher angular resolution 
than that achievable with SPI aboard 
INTEGRAL together with high sensitive
instruments are required in order to 
distinguish and localize - if any - 
discrete source of annihilation 
radiation.

\begin{figure}[!t]
\includegraphics[width=6.3cm]{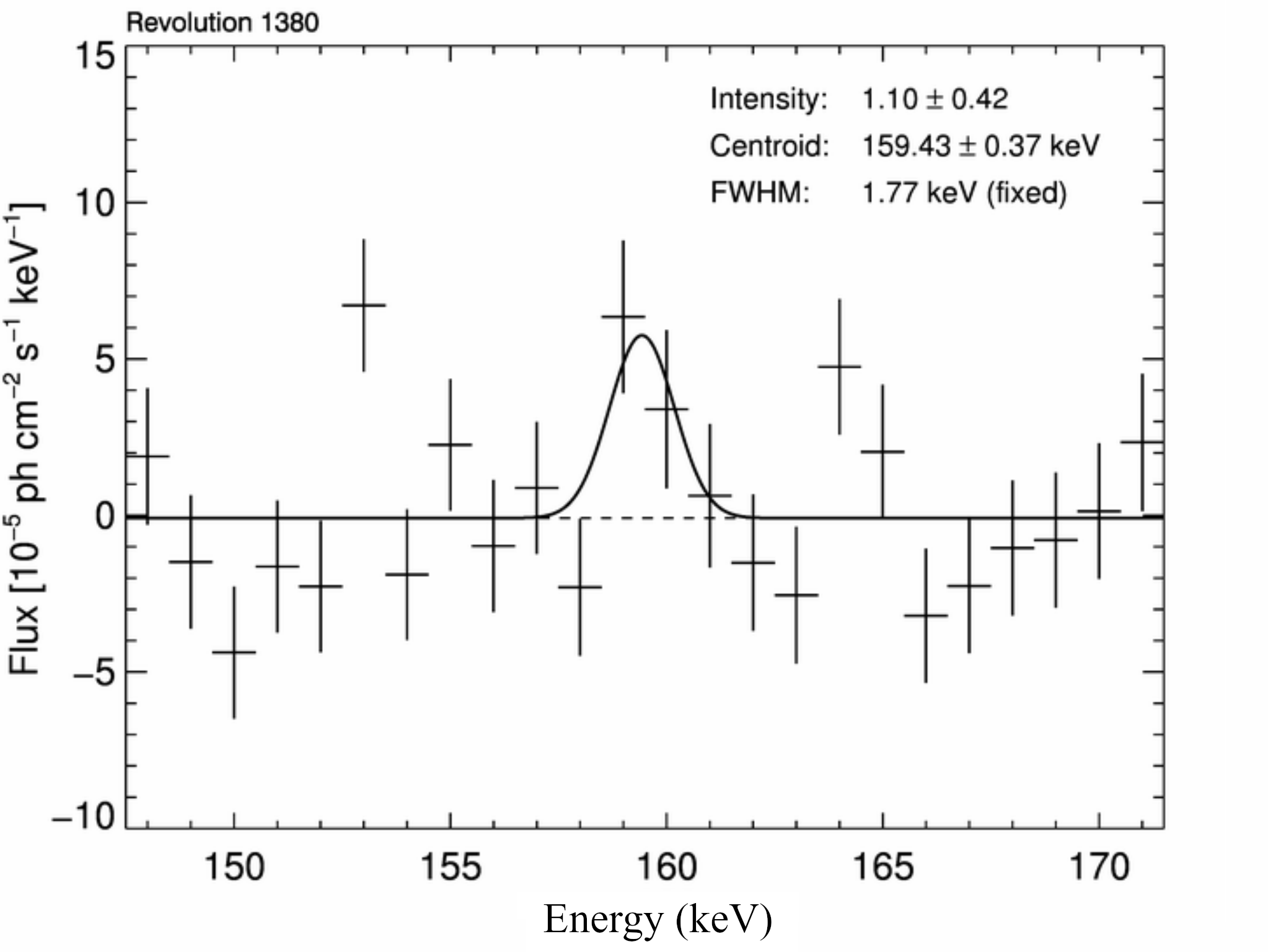}
\includegraphics[width=6.3cm]{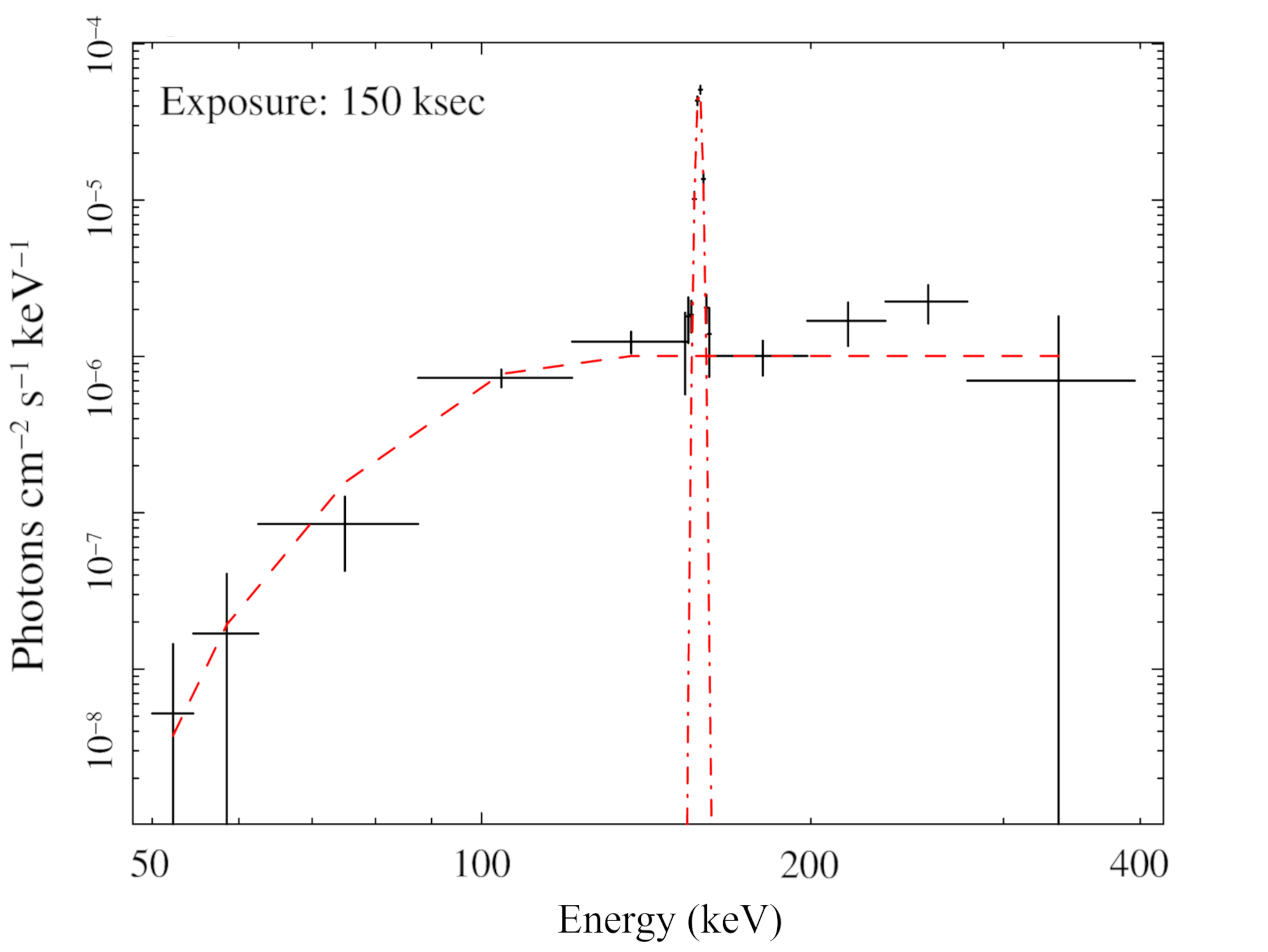}
\caption{Left:  the 158~keV line due to the $^{56}$Ni decay as observed from SN2014J as with the SPI on board INTEGRAL 3 weeks after the explosion. 
Reprinted from~[\!\!\!\citenum{diehl14}]. Right: the same 
line as observed with the NFT onboard 
ASTENA with 150~ks integration time. The 
simulated data (black points) consists
of a continuum modeled with an absorbed power-law 
(red dashed line) with spectral index $\alpha$ = 0 
 normalization $10^{-6}$~ph/cm$^2$/s @ 1~keV plus 
a Gaussian (red dot-dashed line) centered at 158~keV 
with $\sigma$=0.75~keV and normalization 1.1 $\times$ 10$^{-4}$~ph/cm$^2$/s.}
\label{fig:158keV}
\end{figure}

Another topic under study that is worth to 
be mentioned is related to the nucleosynthesis 
of heavy elements in Type-Ia supernovae 
(SNe-Ia). Gamma-rays 
escaping the ejecta of SN-Ia can be used 
as tools for studying both the structure of 
the exploding star and the characteristics of 
the explosion.  One of the key points is to 
estimate  the amount of $^{56}$Ni which is 
probably the most important physical parameter 
underlying the observed correlation of SNe-Ia 
luminosities with their light curves.
The only direct way to estimate this amount is through 
the decay chain $^{56}$Ni $\rightarrow$ 
$^{56}$Co  $\rightarrow$ $^{56}$Fe. 
The decay produces gamma-rays that ultimately power 
the optical light curve of the SN-Ia. 
Most models predict that the ejected 
material is not transparent to gamma-rays 
at least for tens of days since the explosion.
As the ejecta proceeds, it becomes gradually more transparent to gamma-rays.
Contrarily to the expectations, only two weeks 
after SN2014J through SPI/INTEGRAL 
observations, the characteristic gamma-ray lines 
158~keV and 812~keV from the $^{56}$Ni decay 
have been detected[\!\!\citenum{diehl14}] and later confirmed
at 5$\sigma$ confidence level also with ISGRI/INTEGRAL
\cite{isern16}.  The early observations of gamma-rays 
from $^{56}$Ni support the hypotesis of the 
presence of this element at the star surface.
The intensity of the gamma-ray spectra 
mainly depends on the mass and distribution of 
the nickel mass. High sensitivity 
measurements focused at particular 
narrow lines would shed light on 
the SNe-Ia explosion mechanism and, 
ultimately to the correct distribution and mass of 
$^{56}$Ni.

As an example of performances with ASTENA, in Fig.~\ref{fig:158keV} we show the 158~keV line 
detected with SPI/INTEGRAL from SN2014J
\citep{diehl14} for an exposure  
time of 150~ks. The estimated line intensity 
of (1.1$\pm$ 0.4) $\times$ 10$^{-4}$~ph/cm$^2$/s 
corresponds to a detection confidence level of  
2.5$\sigma$. For comparison, with the 
same integration time, a spectrum 
consisting of a 
Gaussian profile (centroid energy 158~keV, 
$\sigma$ = 0.75~keV) superposed to a continuum 
described by a power-law with spectra slope 
$\alpha$=0 and normalization 
$10^{-6}$~ph/cm$^2$/s @ 1~keV has been 
simulated in order to estimate how it 
could be observed with the NFT 
aboard ASTENA. The same significance 
achieved by SPI in 150~ks 
is obtained in $\sim$1~ks.

\section{Conclusions}

Future missions for hard X-ray astrophysics 
require higher sensitivity than current instrumentation and capability of observing 
the sky in a broad energy pass-band. 
Such features will be crucial 
for solving some of the most important open issues. In addition, hard-X and soft gamma-ray
polarimetry, which are still in their infancy, will 
play a crucial role in combination with spectroscopy, 
timing and imaging, providing information about 
the degree and the direction of polarization of the
incident radiation. 
Particularly in hard X-rays, whose emission 
comes directly from the central engine of the sources,
polarimetry is a powerful tool for 
investigating the magnetic field and the distribution 
of matter around astrophysical objects.
Moreover, today hard X-/gamma-ray astrophysics 
suffers from the absence of devices that allow 
the radiation to be focused, the only method 
through which it is possible
to increase the signal to noise of the observations and to suppress the instrumental background.

The ASTENA mission concept includes two complementary 
instruments through which it is possible to carry out 
both polarimetry and imaging in the hard X-ray regime. 
The WFM-IS is an array of 12 monitors that
can effectively work in an 
unprecedented energy pass-band for a single 
instrument (2~keV - 20~MeV) thanks to the coupling 
of SDDs a and scintillator bars.
Thanks to the adoption of a double scale 
coded mask, the instrument enables localization 
accuracy of about 1~arcmin 
and 5~arcmin in the energy 
pass-bands $<$~30~keV and 30 - 150~keV, respectively.
With this location accuracy, the WFM-IS is in  
complete synergy with the second instrument on 
board, the NFT: for the first time a focusing 
optics based on a Laue lens will be operative in 
a broad pass-band and with a FoV of a 4-5~arcminutes 
and an angular 
resolution of a few tens of arcseconds. 
Thanks to its focusing capability it provides 
an outstanding sensitivity for detecting 
source polarization\cite{moita21} and 
both for the continuum and 
for nuclear lines.
The unprecedented broad pass band and the large 
FoV of the WFM-IS will allow to make surveys 
and to detect  faint transient sources, 
including GRBs. The NFT, with its deep 
sensitivity for nearly on-axis sources, 
will be the ideal tool for followup 
observations of detected transient events. 

In this work we have presented the 
ASTENA mission concept and details on 
both instruments aboard. We have described 
some key science issues that can be tackled with 
ASTENA as a stand-alone experiment and in 
synergy with other experiments, including 
its contribution to the multi-messenger 
astrophysics. We expect that a mission 
like ASTENA will be a breakthrough in general 
astrophysics and in particular for 
answering the questions 
that are still central in high energy 
astrophysics.

\section*{Acknowledgements}
This work has been partially supported with 
the financial contribution from the  ASI-INAF agreement n.~2017-14-H.0 
"Studi per future missioni scientifiche", the ASI-INAF agreement n. 2018-10-H.1-2020 " HERMES Technologic Pathfinder", and the AHEAD EU Horizon 2020 project  (Integrated Activities in the High Energy Astrophysics Domain), grant agreement n. 871158.

\bibliographystyle{ws-procs961x669}
\bibliography{ws-pro-sample.bib}

\end{document}